# Uncertainties inherent in the decomposition of a Transformation

Chandru Iyer,
Techink Industries, C – 42, Phase – II, Noida India 201305.
E-mail: chandru_i@yahoo.com

**Abstract:** This contribution adds to the points on the "indeterminacy of special relativity" made by De Abreu and Guerra. We show that the Lorentz Transformation can be composed by the physical observations made in a frame K of events in a frame K' viz i) objects in K' are moving at a speed v relative to K, ii) distances and time intervals measured by K' are at variance with those measured by K and iii) the concept of simultaneity is different in K' compared to K. The order in which the composition is executed determines the nature of the middle aspect (ii). This essential uncertainty of the theory can be resolved only by a universal synchronicity as discussed in [1] based on the unique frame in which the one way speed of light is constant in all directions.

**1. Introduction:** In a recent paper [1], De Abreu and Guerra have completed a trilogy devoted to a journey into the foundations of special relativity. While fully agreeing with their viewpoints, we present here an analysis that brings into play the non-commutative nature of transformations and how a transformation can be decomposed in two different ways with entirely different physical meanings.

The Lorentz transformation has three physical aspects. One is the relative motion between the frames. The second is the observed differences in lengths and running of clocks. The third relates to the adoption of an appropriate synchronization convention [2]. All these three aspects are physically observable; only the last one is attributed sometimes as an appropriate convention, but follows from the constancy of the speed of light as observed by any inertial frame. Comparing the rulers of K' with those of K, we show that whether rulers of K' contracted or elongated in comparison to those of K, depends on the preferred synchronicity convention. In the two specific decompositions that we have shown, the mathematical and physical form of the first and the third aspects remain unchanged.

**2.0. Discussion on Synchronization Convention.** The procedure of synchronizing spatially separated clocks by assuming the one-way speed of light as constant has been termed as the Einsteinian synchronicity convention [2]. However Reichenbach [3] had argued that this is only a conventional synchronicity and there is no compelling reason to adopt this particular synchronicity.

Ohanian [2] has given a complete review of the debate on the conventions relating to synchronization. He also argues that the dynamical considerations forbid any synchronization other than the Einsteinian one, and if an inertial frame adopts a Reichenbach synchronization, Newton's laws would be violated. However, Martinez [4] and Macdonald [5] are not in complete agreement with Ohanian [2].

Martinez [4] has discussed the origin of the Einsteinian synchronization. He observes that the original German word '*festsetzung*' used by Einstein to prescribe the Einsteinian synchronization has been translated
into English as 'stipulation' and into French as 'convention.' Eddington also advanced the concept that the Michelson-Morley experiment only determined the round trip speed of a light ray as a constant and a synchronization convention was needed to further specify that the speed of light remained constant on both the onward and return trips [4]. Macdonald argues that Einstein definitely intended the synchronization proposed by him as a method or definition. And this is the reason Einstein emphasized that his definition "is in reality neither a *supposition nor a hypothesis* about the physical nature of light, but a stipulation (*festsetzung*) which I can make of my own free will in order to arrive at a definition of simultaneity"[5].

In his reply to the comments by Martinez [4] and Macdonald [5] Ohanian [6] has argued that when the Einsteinian synchronization convention is adopted in all inertial reference frames, "it permits us to express the laws of physics in their simplest from." He further states that "The adoption of a preferential inertial reference frame in which all the laws of physics take their simplest from compels the E (Einsteinian ) synchronization and forbids the R (Reichenbach) synchronization" [6] .

Thus the synchronization adopted as a consequence of the second postulate is a convention or an "internal synchronization" as discussed by De Abreau and Guerra [1, pp37], that makes the inertial frame conform to the second postulate viz constancy of the speed of light as observed by all inertial frames.

### 3.0. Decomposition of Lorentz Transformation:

The Lorentz transformation may be constructed as a composition of i) Galilean motion, ii) Lorentz length contraction – time dilation and iii) frame specific re-synchronization.

Galilean motion in uni-dimensional space is represented by the transformation of event coordinates.

$$A = \begin{pmatrix} 1 & -v \\ 0 & 1 \end{pmatrix} \quad \ldots\ldots\ldots\ldots\ldots\ldots\ldots\ldots\ldots\ldots\ldots\ldots\ldots(1)$$

The above transformation of event coordinates is from frame K to frame K', moving at v with respect to K, in the Galilean framework. In the relativistic framework, it can be construed as giving the two parameters (as observed by K), the time at which the event occurred and the distance between the location of the event and the location of the origin of K' at the same instant at which the event occurred in K as determined by the Einstenian synchronicity convention of inertial frame K.

The event coordinates after applying the transformation A as above, will give the distance and time coordinates of the event where i) A clock, co-moving with K and present at the space-time location of the event directly measures the time coordinate $t^*$ , ii) the position P, as observed by K, of the origin of K' at t= $t^*$ is identified by using the Einstenian synchronization convention [2] as applicable to K and iii) the distance between P and the location of the event is measured as $x^*$. The coordinates ($x^*$, $t^*$) give a set of unique coordinates for the event.

The Lorentz length contraction – time dilation is represented by –

$$B = \begin{pmatrix} \gamma & 0 \\ 0 & 1/\gamma \end{pmatrix} \quad \text{...........................(2)}$$

Where $\gamma = 1/\sqrt{1-(v^2/c^2)}$

After applying the transformation BA, we get the coordinates of the event as observed by K' but still using the Einstenian synchronization convention of K. This transformation BA is also the IST transformation as discussed in [1, pp 36].

And the Lorentz asynchronization is represented by

$$D = \begin{pmatrix} 1 & 0 \\ \frac{-v}{c^2} & 1 \end{pmatrix} \quad \text{............................ (3)}$$

Where the asynchronization term is- $\frac{-v(x-vt)\gamma}{c^2} = \frac{-vx'}{c^2}$

When we apply D to BA, it transports the synchronization convention from Einstenian K to Einstenian K'.

Thus the transformation DBA fully transports the event coordinates observed by an observer situated at the origin of K, using Einstenian synchronicity of K and co-moving with K to that observed by an observer situated at the origin of K', using Einstenian synchronicity of K' and co-moving with K'. DBA is the Lorentz transformation. The three components may be termed as i) movement of objects of K' represented by A, ii) contraction of objects in K' along line of motion and slow running of clocks of K' represented by B and iii) switching from Einstenian synchronization convention of K to that of K' represented by D. And these three are applied in that order and the Lorentz transformation is composed as DBA.

The Lorentz transformation can also be composed as i) first switching from the Einstenian synchronization convention of K to that of Einstenian synchronization convention of K'. ii) elongation of the rods of K' and faster running of clocks of K' and iii) movement of objects of K'. This will be represented by A $B^{-1}$ D. In this case, we observe that the clocks of K are asynchronized by + $vx/c^2$ about the origin of K, when compared with those of K'. In order to correct this, we apply the transformation D and that 'corrects' the asynchronization of K with respect to K' and switches the synchronicity convention adopted from Einstenian synchronicity convention of K to Einstenian synchronicity convention of K' ; ii) then it is observed that the rulers in K' are elongated and the clocks of K' are running faster. To account for this, we apply the transformation $B^{-1}$. iii) then we account for the relative motion by the transformation A, which accounts for the motion of K' at +v with respect to K. Thus we obtain the transformation of event coordinates from K to K' as $AB^{-1}D$.

In both these decompositions, we were transforming from K to K'. Only in the first case we switched the synchronization convention at the end and in the second case we switched the synchronization convention at the outset. In the first case we obtained the Lorentz transformation as DBA and in the second case we obtained the Lorentz transformation as $AB^{-1}D$. The physical aspects represented by matrices A and D remain unchanged; the physical aspect represented by B becomes inversed when the order of the composition is reversed. The important aspect to note is that both the compositions lead to the Lorentz transformation.

$$DBA = A\ B^{-1}\ D \quad \text{..........................................................(4)}$$

From a mathematical stand point, the rearranging of the order of the matrices and inversing only one of them, leaves the overall transformation unaltered. Physically, it means the 'timing' of the switch in the synchronization convention determines whether the rods contracted or elongated. What this means is that, if we account for movement, ruler deformations, the way clocks run and switching of synchronization convention, in that order, we find that the rulers contract and clocks run slow. But when we switch the synchronization convention first and then account for the rest of the physical effects, we observe that the rulers have expanded and the clocks are running fast.

If K is presumed to be at 'rest' then the Einstenian synchronization of K is the correct synchronization. In this case when transforming the event coordinates from K to K' we switch to the Einstenian synchronization of K' at the end only, as it is only the convention of K' and not the actual synchronicity. In this case, we observe that the rulers in K' have contracted and the clocks are running slow.

If K' is presumed to be at rest, then, when transforming the event coordinates from K to K' we switch to the Einsteinian synchronicity of K' first as this is the correct synchronicity to observe physical processes. In this case, we observe that the rulers have expanded and the clocks are running faster in K'.

When we take the inverse of the Lorentz transformation represented as DBA as in equation (4), the inverse is $A^{-1} B^{-1} D^{-1}$

Where $D^{-1} = \begin{pmatrix} 1 & 0 \\ \frac{v}{c^2} & 1 \end{pmatrix}$ ; $B^{-1} = \begin{pmatrix} \frac{1}{\gamma} & 0 \\ 0 & \gamma \end{pmatrix}$ ; $A^{-1} = \begin{pmatrix} 1 & +v \\ 0 & 1 \end{pmatrix}$

By the same principle represented by equation (4),

We have $A^{-1} B^{-1} D^{-1} = D^{-1} B A^{-1}$ ........................ .................... (5)

The inverse of the Lorentz transformation $(DBA)^{-1}$ represented on the left hand side of equation (5) when viewed as expressed in right hand side of equation (5) appears exactly like the Lorentz transformation with only the sign of the velocity reversed.

**4.0. Summary:** The Lorentz Transformation of event coordinates as observed by K to K' can be composed by accounting for the observed physical phenomena viz i) relative motion ii) differences in observed distances and time intervals, iii) change in synchronization convention. This can be done in two ways.

$$\begin{pmatrix} \gamma & -v\gamma \\ \frac{-v\gamma}{c^2} & \gamma \end{pmatrix} = \underbrace{\begin{pmatrix} 1 & 0 \\ \frac{-v}{c^2} & 1 \end{pmatrix} \begin{pmatrix} \gamma & 0 \\ 0 & \frac{1}{\gamma} \end{pmatrix}}_{\text{IST transformation [1]}} \begin{pmatrix} 1 & -v \\ 0 & 1 \end{pmatrix}$$

Change in Synchronization Convention | Contraction of rulers and slow running of clocks. | Relative motion

$$\begin{pmatrix} \gamma & -v\gamma \\ \frac{-v\gamma}{c^2} & \gamma \end{pmatrix} = \begin{pmatrix} 1 & -v \\ 0 & 1 \end{pmatrix} \begin{pmatrix} \frac{1}{\gamma} & 0 \\ 0 & \gamma \end{pmatrix} \begin{pmatrix} 1 & 0 \\ \frac{-v}{c^2} & 1 \end{pmatrix}$$

Relative motion | Elongation of rulers and faster running of clocks. | Change in synchronization convention

The change in the middle matrix describes the indeterminacy as described in [1]. The other two matrices remain the same. Normally the composition is shown as the former, but there is no reason the composition cannot take the later form.

**5.0. Conclusion:** The Lorentz Transformation contains the mathematical formulations for transforming the event coordinates of any arbitrary event from the ones observed by inertial frame K to those observed by inertial frame K', where K' is moving at velocity v with respect to K. The observation of K about K' itself are i) Objects in K' are moving at v. ii) The measurement of distance and time intervals by K' are at variance with those of K and iii) The concept of synchronicity of K' is different from that of K. So it may be reasonable to compose the Lorentz Transformation as a composition of these three observations of K about K'. We have shown that if the composition is executed in the order as above (DBA), then the second step is contraction of rulers and slow running of clocks; if the composition is executed in the reverse order (AB$^{-1}$D), then the second step is elongation of rods and faster running of clocks. This essentially summarizes the "indeterminacy of special relativity" as detailed in [1]. The indeterminacy is contained in the uncertainty whether the rulers contracted or expanded and the clocks run slower or faster in K' compared to those of K. Only a universal synchronicity convention as the one described in [1] using the unique frame wherein the one-way speed of light is constant (in all directions), can resolve this uncertainty.

To generalize the above observations, one may state that to express the composition of a physical transformation as a combination of discrete processes is a natural endeavor of Physics. When such a composition is non-commutative and offers different alternatives, a preferred order of composition is implicit.

**Acknowledgements**: The article has benefited immensely by the comments of an anonymous referee.

**References:**
[1] De Abreu and Guerra. EJP 29 (2008) pp33 – 52. The principle of relativity and the indeterminacy of special relativity.

[2] Ohanian, H The role of dynamics in the synchronization problem. *Am. J. of Phys.* **72** *(2004) (2) pp 141 – 148.*

*[3]* Reichenbach, H. (1958). The Philosophy of Space and Time. Dover, New York, p.127.

*[4] Martinez, A Conventional and inertial reference frames Am J.Phys.(5). pp 452 – 454.*

[5] Macdonald, A Comment on' the role of dynamics in the synchronization problem', by H.Ohanian Am. J. Phys. 72(2) 141-148. 2004. Am. J. Phys. 73(5) pp 454 – 455.

[6] Ohanian, H. Reply to comment (s) on 'the role of dynamics in the synchronization problem' by A Macdonald, [Am. J. Phys. 73, 454] and A. Martinez [Am. J.Phys. 73, 452.] Am. J. phys. 73 (5) , pp 456 – 457.

Appendix: The speed of light under the IST and Lorentz Transformations

The IST transformation is given in one-dimensional space as

x' = (x-vt)$\gamma$ ; t' = t/$\gamma$

Where the relative motion is along the x, x' axis and $\gamma = 1/\sqrt{1-(v^2/c^2)}$

In three-dimensional space the two additional equations y' = y; z' = z are in order.

Expressed in the matrix notation the transformation becomes.

$$\begin{pmatrix} x' \\ y' \\ z' \\ t' \end{pmatrix} = \begin{pmatrix} \gamma & 0 & 0 & -v\gamma \\ 0 & 1 & 0 & 0 \\ 0 & 0 & 1 & 0 \\ 0 & 0 & 0 & 1/\gamma \end{pmatrix} \begin{pmatrix} x \\ y \\ z \\ t \end{pmatrix} \qquad \ldots\ldots\ldots\ldots(6)$$

Where ( x, y, z, t) are the event coordinates in frame K and x', y', z', t', are the event coordinates in K'.

For a light ray observed by K, to be propagating along an arbitrary line with angle $\phi$ with z-axis and with its projection on the xy plane subtending an angle $\theta$ with the x axis,

We have the following event generated at any time t (as observed by K).

$x = ct \sin\phi \cos\theta$

$y = ct \sin\phi \sin\theta$

$z = ct \cos\phi$

The above event is transformed by the IST transformation to

$$\begin{pmatrix} \gamma & 0 & 0 & -v\gamma \\ 0 & 1 & 0 & 0 \\ 0 & 0 & 1 & 0 \\ 0 & 0 & 0 & 1/\gamma \end{pmatrix} \begin{pmatrix} ct\sin\phi\cos\theta \\ ct\sin\phi\sin\theta \\ ct\cos\phi \\ t \end{pmatrix} = \begin{pmatrix} x' \\ y' \\ z' \\ t' \end{pmatrix}$$

Noting t' = t/$\gamma$, we get

$$\begin{pmatrix} x' \\ y' \\ z' \\ t' \end{pmatrix} = \begin{pmatrix} c\gamma^2 t' \sin\phi\cos\theta - v\gamma^2 t' \\ c\gamma t' \sin\phi\sin\theta \\ c\gamma t' \cos\phi \\ t' \end{pmatrix} \quad \ldots\ldots\ldots\ldots(7)$$

The components of the observed velocity become

$$C_{x'}^+ = c\gamma^2 \sin\phi\cos\theta - v\gamma^2 \quad \ldots\ldots\ldots\ldots\ldots(8)$$

$$C_{y'}^+ = c\gamma \sin\phi\sin\theta \quad \ldots\ldots\ldots\ldots\ldots(9)$$

$$C_{z'}^+ = c\gamma \cos\phi \quad \ldots\ldots\ldots\ldots\ldots(10)$$

Squaring and adding the component velocities and then taking the square root, we get

$$C^+ = c\gamma \sqrt{\gamma^2 \left(\sin\phi\cos\theta - \frac{v}{c}\right)^2 + \sin^2\phi\sin^2\theta + \cos^2\phi}$$

Substituting for $\gamma^2$ as $\dfrac{1}{1 - v^2/c^2}$ and simplifying we get

$$C^+ = c\gamma \sqrt{\frac{\left(\sin\phi\cos\theta - \dfrac{v}{c}\right)^2}{1 - \dfrac{v^2}{c^2}} + \sin^2\phi\sin^2\theta + \cos^2\phi}$$

$$= c\gamma^2 \sqrt{\left(\sin\phi\cos\theta - \frac{v}{c}\right)^2 + \left(\sin^2\phi\sin^2\theta + \cos^2\phi\right)\left(1 - \frac{v^2}{c^2}\right)}$$

Simplifying the expression under square root, we get

$$C^+ = c\gamma^2 \left(1 - \frac{v}{c}\sin\phi\cos\theta\right) \quad \ldots\ldots\ldots\ldots\ldots(11)$$

The above formula for the observed speed of light is in terms of the direction cosines observed by K. To convert the same in terms of direction cosines observed by K', we proceed as below.

Dividing both sides of equations (8), (9) and (10) by corresponding sides of equation (11), we get

$$\cos\theta' \sin\phi' = \frac{\sin\phi\cos\theta - \frac{v}{c}}{1 - \frac{v}{c}\sin\phi\cos\theta} \qquad \ldots\ldots\ldots\ldots\ldots\ldots(12)$$

$$\sin\theta'\sin\phi' = \frac{\sin\theta\sin\phi}{\gamma\left(1 - \frac{v}{c}\sin\phi\cos\theta\right)} \qquad \ldots\ldots\ldots\ldots\ldots\ldots(13)$$

And

$$\cos\phi' = \frac{\cos\phi}{\gamma\left(1 - \frac{v}{c}\sin\phi\cos\theta\right)} \qquad \ldots\ldots\ldots\ldots\ldots\ldots(14)$$

One may note that given $\theta$ and $\phi$, $\phi'$ can be evaluated from (14) and $\theta'$ can be evaluated from either (12) or (13).

From equation (12) we obtain (by treating ($\sin\phi\cos\theta$) as a single variable and solving for the same)

$$\sin\phi\cos\theta = \frac{\sin\phi'\cos\theta' + \frac{v}{c}}{1 + \frac{v}{c}\sin\phi'\cos\theta'} \qquad \ldots\ldots\ldots\ldots\ldots\ldots(15)$$

Substituting for $\sin\phi\cos\theta$ from equation (15) into equation (11), we obtain

$$C^+ = \frac{c}{1 + \frac{v}{c}\sin\phi'\cos\theta'} \qquad \ldots\ldots\ldots\ldots\ldots\ldots(16)$$

Equation (16) expresses the observed velocity of light in K' as a function of the observed direction cosines of the line of propagation in K'.

And for any distance $\Delta L'$ in any arbitrary direction denoted by $\phi'$ and $\theta'$, the time taken will be

$$\Delta t' = \frac{\Delta L'}{C^+}$$

Substituting for $C^+$ form equation (16), we obtain

$$\Delta t' = \frac{\Delta L'}{c}\left(1 + \frac{v}{c}\sin\phi'\cos\theta'\right)$$

$$= \frac{\Delta L'}{c} + \frac{v}{c^2}\Delta L'\sin\phi'\cos\theta'$$

Noting that $\Delta L'\sin\phi'\cos\theta' = \Delta x'$, we obtain

$$\Delta t' = \frac{\Delta L'}{c} + \frac{v}{c^2}\Delta x' \quad \ldots\ldots\ldots\ldots\ldots(17)$$

In any closed path the summation of the second term vanishes and thus the average round trip speed of light is observed to be c.

Further when we shift to the "proper" synchronization convention of K' given by

$$t'_E = t' - \frac{vx'}{c^2} \quad \ldots\ldots\ldots\ldots\ldots(18)$$

or $\Delta t'_E = \Delta t' - \frac{v}{c^2}\Delta x'$

or $\quad \Delta t' = \Delta t'_E + \frac{v}{c^2}\Delta x' \quad \ldots\ldots\ldots\ldots\ldots(19)$

Comparing (17) and (19), we get

$$\Delta t'_E = \frac{\Delta L'}{c} \quad \ldots\ldots\ldots\ldots\ldots(20)$$

Therefore under the proper synchronization convention of K', not only the round trip speed but also the one-way speed of light remains constant.

Conclusion: Under the IST transformation, the average speed of light in any closed path is constant. In any given line segment, the speed is given by equation (16). The Lorentz transformation is obtained from the IST transformation by switching to the 'proper' synchronization convention of the target inertial frame, K'. Under the Lorentz transformation, the speed of light remains constant in every segment of the path.